\documentstyle{elsart}
\input epsf
\begin{document}
\begin{frontmatter}
\title{Log-periodic self-similarity: an emerging financial law?}
\author[IFJ,KFA,URZ,UBONN]{S. Dro\.zd\.z},
\author[KFA]{F. Gr\"ummer},
\author[WLB]{F. Ruf},
\author[KFA]{J. Speth}

\address[IFJ]{Institute of Nuclear Physics, Radzikowskiego 152,
31--342 Krak\'ow, Poland}
\address[KFA]{Institute f\"ur Kernphysik, Forschungszentrum J\"ulich,
D--52425 J\"ulich, Germany}
\address[URZ]{Institute of Physics, University of Rzesz\'ow,
PL-35-310 Rzesz\'ow, Poland}
\address[UBONN]{Physikalisches Institut, Universit\"at Bonn, D-53115 
Bonn, Germany}
\address[WLB]{West LB International S.A., 32--34 bd
Grande--Duchesse Charlotte, L--2014 Luxembourg}

\begin{abstract}
A hypothesis that the financial log-periodicity, cascading self-similarity 
through various time scales, carries signatures of a law is pursued.
It is shown that the most significant historical financial events can
be classified amazingly well using a single and unique value of the
preferred scaling factor $\lambda=2$, which indicates that its real value   
should be close to this number. This applies even to a declining   
decelerating log-periodic phase. Crucial in this connection is      
identification of a ``super-bubble'' (bubble on bubble) phenomenon.
Identifying a potential ``universal''
preferred scaling factor, as undertaken here, may significantly improve   
the predictive power of the corresponding methodology.
Several more specific related results include evidence that:

\noindent (i) the real end of the high technology bubble on the 
stock market started (with a decelerating log-periodic draw down) 
in the begining of September 2000;

\noindent (ii) a parallel 2000-2002 decline
seen in the Standard $\&$ Poor's 500 from the log-periodic perspective
is already of the same significance as the one
of the early 1930s and of the late 1970s;

\noindent (iii) all this points to a much more serious global
crash in around 2025, of course from a level much
higher (at least one order of magnitude) than in 2000. 
\end{abstract}   

\end{frontmatter}
 
{\it e-mail: Stanislaw.Drozdz@ifj.edu.pl}

\bigskip
 
\section{Introduction}

The suggestion that financial dynamics may be governed by phenomena
analogous to criticality in the statistical physics sense and,
especially, the related subtle concept of
log-periodicity~\cite{Sorn1,Feig1,Vand,Drozdz} proves exciting and at the
same time somewhat controversial~\cite{Lalo,Ilin,Feig2}. In its
conventional form criticality implies a scale invariance which, for a
properly defined function $F(x)$ characterizing the system, means that
\begin{equation}
F(\lambda x) = \gamma F(x).
\label{eq:F}
\end{equation}
A positive constant $\gamma$ in this equation describes how the properties
of the system change when it is rescaled by the factor $\lambda$. One
obvious solution to this equation is:
\begin{equation}
F_{0}(x) = x^{\alpha},
\label{eq:pow}
\end{equation}
where $\alpha = \ln(\gamma)/\ln(\lambda)$. It represents a standard
power-law that is characteristic of continuous scale-invariance and
$\alpha$ is the corresponding critical exponent.

The zig-zag character of financial dynamics attracts attention to the
general solution~\cite{Naue} of Eq.~(\ref{eq:F}):
\begin{equation}
F(x) = x^{\alpha} P({\ln (x) / \ln(\lambda)}).
\label{eq:logper}
\end{equation}

$P$ denotes a periodic function of period one. The dominating scaling
(\ref{eq:pow}) thus acquires a correction that is periodic in $\ln(x)$.
This solution can be interpreted in terms of discrete
scale-invariance~\cite{Sorn2} and a complex critical exponent~\cite{Newm}.
A functional form of $P$ is not determined at this level. It only demands
that if
\begin{equation}
x = \vert T - T_c \vert,
\label{eq:x}
\end{equation}
where T denotes the ordinary time labeling the original price time series,
represents a distance to the critical point $T_c$, the resulting spacings
between the corresponding consecutive repeatable structures at $x_n$ (i.e.,
minima or maxima) of the log-periodic oscillations seen in the linear
scale follow a geometric contraction according to the relation
\begin{equation}
{x_{n+1} - x_n \over x_{n+2} - x_{n+1}} = \lambda.
\label{eq:gp}
\end{equation}
The critical points coincide with the accumulation of such oscillations
and, in the context of the financial dynamics, it is this effect that
potentially can be used for prediction provided $\lambda$ is really well
defined and constant. Our previous contribution~\cite{Drozdz} provides two
related elements that turn out to be essential for a proper interpretation and
handling of the financial patterns. One is the suggestion that consistency of
the theory requires that, if applicable, the log-periodic scenario is to
manifest its action self-similarly through various time scales. Imprints
of such effects have also been found~\cite{Drozdz} in the real stock
markets and further confirmed in Ref.~\cite{Johansen}. Second is
identification~\cite{Drozdz} that $\lambda \approx 2$ is the most
appropriate preferred scaling factor through various time scales, in
amazing consistency with those found for a whole variety of other complex
systems~\cite{Sorn2,Newm,Ball,Sale}. Below we present an attempt to
classify all the significant historical events on the world's leading stock
market, including the 2000--2002 declining and log-periodically
decelerating phase, within such a scheme.

\begin{figure}
\hspace{0.5cm}
\epsfxsize 12.0cm
\epsffile{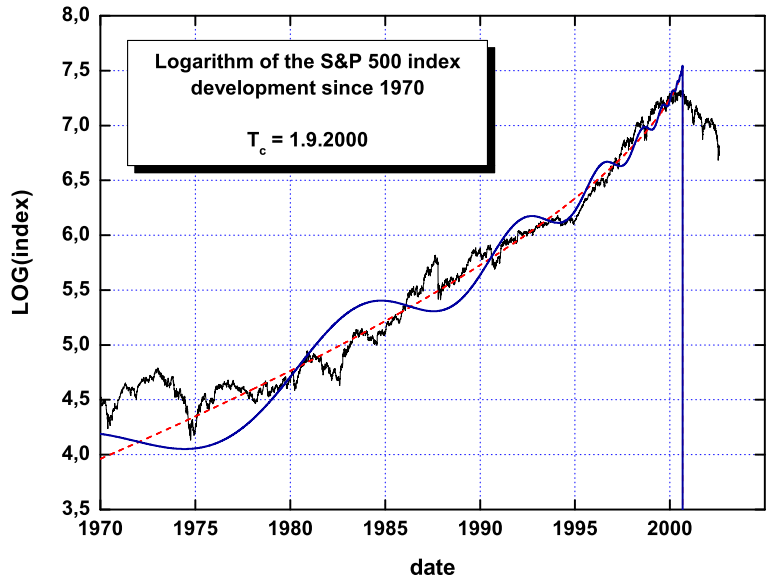}

\vspace{-0.5cm}
\hspace{0.5cm}   
\epsfxsize 12.0cm   
\epsffile{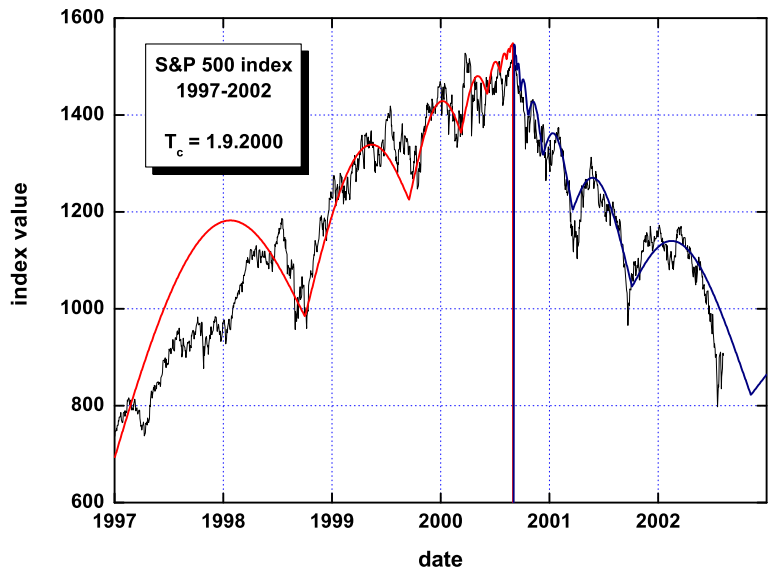}
\caption{(a) Logarithm of the Standard $\&$ Poor's
500 over the period 1970--2002 versus its corresponding log-periodic
representation (solid line) in terms of Eqs.~(\ref{eq:logper}) and
(\ref{eq:pfo}). (b) S$\&$P500 from 1997 till end of August 2002. The solid
lines serve as a reference illustrating the log-periodically accelerating
and decelerating scenarios with a common time of crash $T_c = 1.9.2000$.
The modulus of the $cosine$ in Eq.~(\ref{eq:pfo}) is found here to
constitute a better representation. The preferred scaling factor $\lambda
= 2$ is used everywhere.}
\label{fig:1}
\end{figure}

\section{Log-periodic S$\&$P500 in 1970 - 2002}

The above period includes essentially the whole spectrum of effects of
interest from the present perspective. It seems that the best scalar
representation of the world global economic development during this period
is in terms of the Standard $\&$ Poor's 500 index. Keeping in mind that there
exists some freedom in choosing a specific form of the periodic function
$P$ in Eq.(\ref{eq:logper}), which imposes a serious restriction on the
mathematical rigour of the corresponding methodology, we take the first
term of its Fourier expansion,
\begin{equation}
P(\ln(x)/\ln(\lambda)) = A + B \cos ({\omega \over 2\pi} \ln(x) + \phi).
\label{eq:pfo}
\end{equation}
This of course implies that $\omega = 2\pi / \ln(\lambda)$. A unit used
to measure $x$ (equivalently $T$) can be absorbed into $\phi$. With our
$\lambda = 2$ we then try to obtain the best representation of the
oscillatory structure seem in the real market during this period. The
other parameters of Eqs.~(\ref{eq:logper}) and (\ref{eq:pfo}) are not
relevant for the present discussion and, since they are also
nonuniversal~\cite{Drozdz}, are not listed here. The result for the
S$\&$P500 is shown in Fig.~\ref{fig:1}a and can be seen to remain
systematically in phase with the corresponding market trends approximately
pointing to September 1, 2000 as the date of the reversal of the 
almost 20 years global increasing trend. It is interesting to see that the famous
Black Monday of October 19, 1987, fits perfectly and constitutes one of
the prominent log-periodic precursors of a more serious global crash that
started in September 2000. A closer inspection of the vicinity of this
date, obtained by the magnification presented in Fig.~\ref{fig:1}b, shows
two other relevant elements. It clearly indicates that the modulus of
the $cosine$ in Eq.~(\ref{eq:pfo}) provides a better representation for the
log-periodic modulation. Secondly, it provides independent evidence
that the real date $T_c$ marking reverse of the upward global trend is
the begining of September 2000, as it is exactly at this time that the
decelerating log-periodic oscillations accompanying the decline start. Such an
impressive synchronization of the end of a log-periodically accelerating
bubble phase with the begining of log-periodically decelerating
``anti-bubble''~\cite{Johansen1} phase is spectacular, and can indeed
be considered as an extra argument in favour of the consistency of the 
log-periodic scenario may offer. 
Furthermore, the same preferred scaling factor
$\lambda = 2$ has been used in the analytical representation for the
decelerating phase shown in Fig.~\ref{fig:1}b, and it looks optimal.

\begin{figure}
\hspace{0.5cm}
\epsfxsize 13.0cm
\epsffile{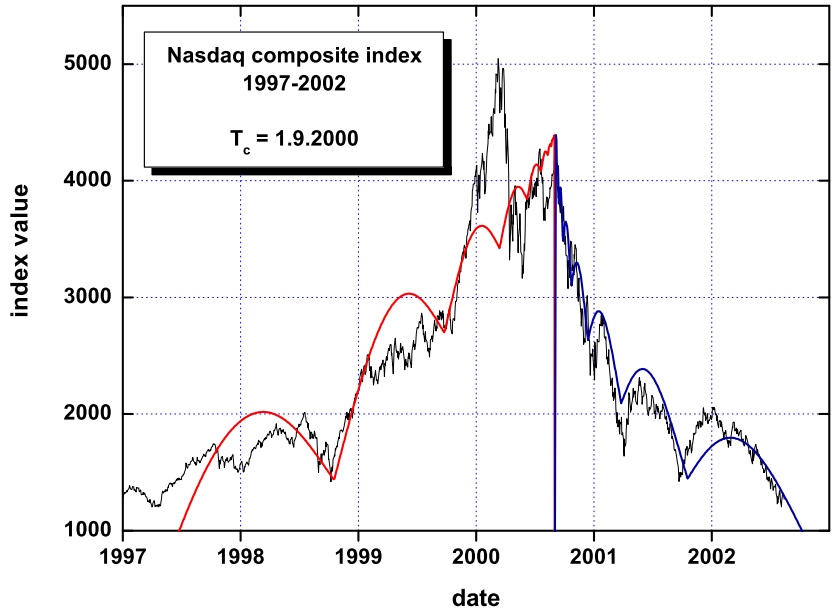}

\vspace{-0.5cm}
\hspace{0.5cm}   
\epsfxsize 13.0cm   
\epsffile{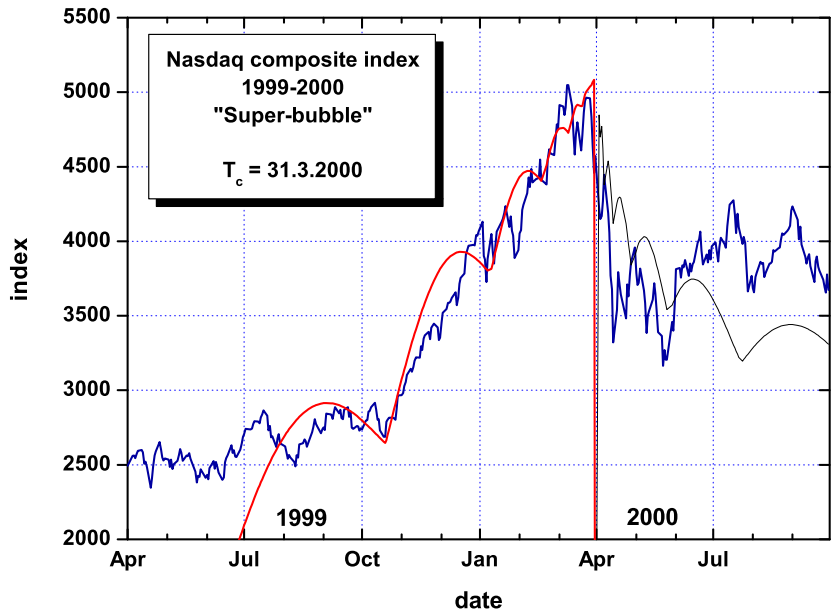}
\caption{(a) The Nasdaq Composite from 1997 through
August 2002 and its log-periodic representation (solid line) in terms of
Eqs.~(\ref{eq:logper}) and (\ref{eq:pfo}), with the modulus of the
$cosine$. (b) The Nasdaq ``super-bubble'' of November 1999 -- March 2000.
The solid line illustrates its own short term log-periodic development.
The preferred scaling factor $\lambda = 2$ is used throughout.}
\label{fig:2}
\end{figure}

\section{Phenomenon of a ``super-bubble''}

The last few years of the stock market development during the period
discussed above was driven by the high-technology sector, whose
appropriate measure is provided by the Nasdaq. How its specific
time-dependence relates to the S$\&$P500 of Fig.~\ref{fig:1}b, especially
in the context of the log-periodic phase transition seen there, is thus a
natural and intriguing question. Since the high technology sector has been
the leader in dictating the global trend, one expects the same scenario to
apply. While this is true during the decline starting in September 2000,
as can be seen from Fig.~\ref{fig:2}a, the Nasdaq development does not,
however, parallel exactly that of the S$\&$P500 during the bubble
phase. The Nasdaq value in March 2000 is significantly higher than at
$T_c$ of September 1. This may tempt one to view~\cite{Johansen2} late
March 2000 as the time marking the end of the high-technology speculative
bubble. This, to some extent, may be considered a matter of taste, although
the Nasdaq clearly follows its long term trend precisely until September.
Its oscillation patterns in time also coincide with accumulation of
oscillations pointing to September, as prescribed by our postulated
universal value of $\lambda$. Reconcilation within the spirit of the
hierarchy of self-similar log-periodic patterns can be obtained by the
following additional postulate, which allows one to better understand the
subtleties of the underlying dynamics: The substructure in the period
November 1999 - March 2000, as one of the consecutive increases in the
sequence of long term log-periodic pattern, gets boosted into a local
bubble on top of a long term bubble, and therefore we term it a
``super-bubble''. This local ``super-bubble'' then crashes (as at the
end of March 2000) and the system returns to a normal bubble state that
eventually crashes at the time (here September 2000) determined by the
long term patterns. In fact, a trace of the same frequency log-periodic
oscillations can even be seen to accompany the dynamics of this
``super-bubble'', as shown in Fig.~\ref{fig:2}b.

\begin{figure}
\hspace{0.5cm}
\epsfxsize 13.0cm
\epsffile{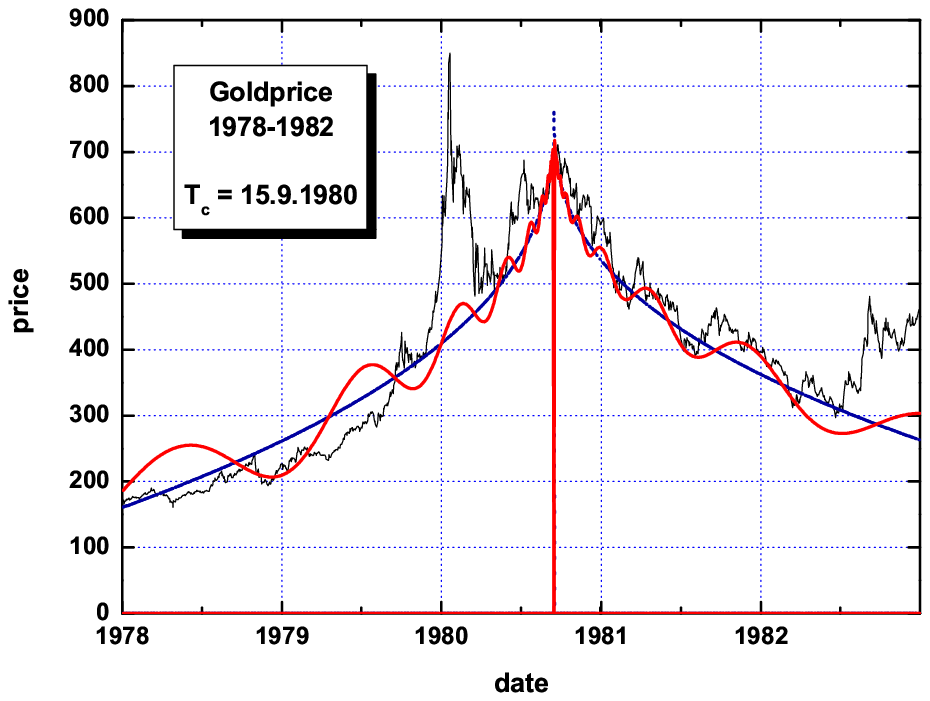}

\vspace{-0.5cm}
\hspace{0.5cm}
\epsfxsize 13.0cm
\epsffile{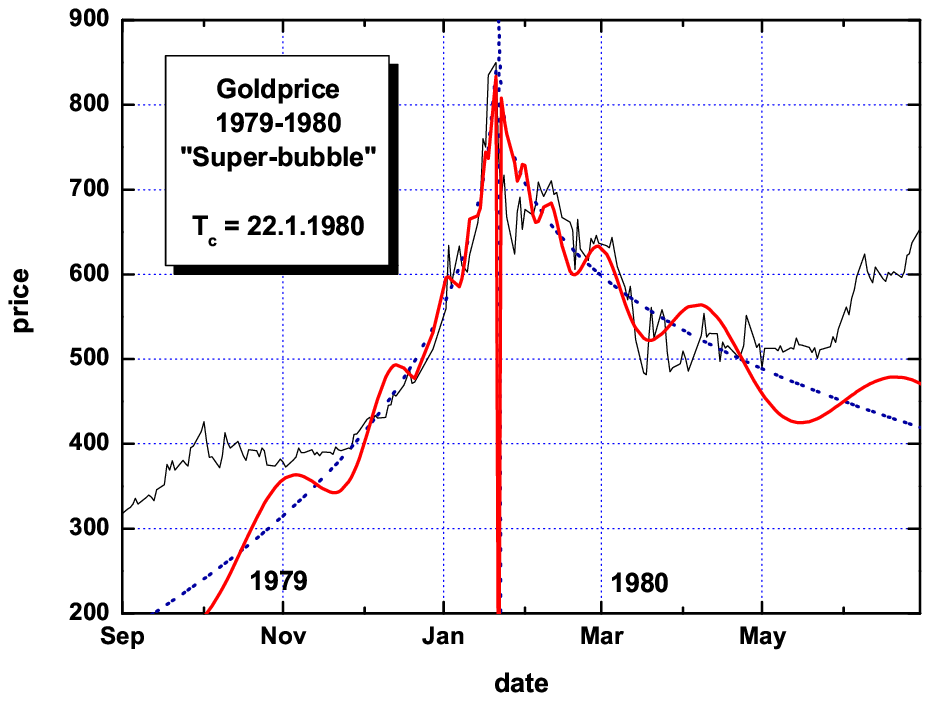}
\caption{The gold price over the period 1978--1982
and the corresponding log-periodic representations in terms of
Eqs.~(\ref{eq:logper}) and (\ref{eq:pfo}), on both sides of $T_c =
15.9.1980$ with the same $\lambda = 2$. (b) The gold price ``super-bubble''
compared to the optimal short term log-periodic scenarios with the same
preferred scaling factor.}
\label{fig:3}
\end{figure}

As far as such exotic effects are concerned, the Nasdaq is not an
exception. In this regard even more spectacular, as illustrated in
Fig.~\ref{fig:3}a and Fig.~\ref{fig:3}b, was the gold price development
around 1980. Its extremely sharp log-periodic $(\lambda = 2)$ bubble also
boosts one of its upgoing substructures into a ``super-bubble'', which
itself develops its own log-periodic oscillations with the same scaling
factor, and the global long term log-periodic bubble eventually starts
decaying, also log-periodically with $\lambda = 2$. Such a scenario
also resolves the difficulty encountered in Ref.~\cite{Johansen2} of
filling the gap between the gold price maximum and the onset of the
decelerating log-periodic phase. Our general remark in this context is that
overlooking such effects of the ``super-bubbles'' may lead to a whole
spectrum of $\lambda$'s, which is both unaesthetic and misleading.

\section{Looking into future}

\begin{figure}
\hspace{0.5cm}
\epsfxsize 13.0cm
\epsffile{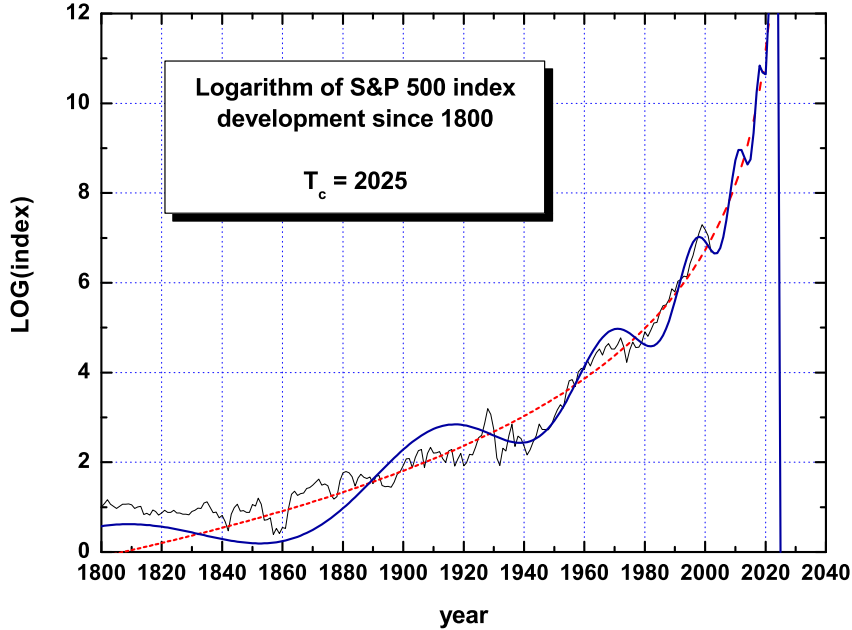}
\caption{Logarithm of the Standard $\&$ Poor's 500 index
illustrating its development since 1800~\cite{globalfin}.
The values prior to its official introduction are reconstructed 
from historical data~\cite{globalfin}.
The solid line represents the
optimal log-periodic representation appropriate to this time scale with
the usual preferred scaling factor. The fact that the third minimum in the
solid line comes a little bit later than in the real data may seem somewhat
disturbing. In this connection, however, it is interesting to notice that
when correcting for a huge inflation, especially at that time, the minimum
in the real data gets shifted to the early 1980s.}
\label{fig:4}
\end{figure}

Having collected from several time scales quite interesting evidence of
universality of the financial log-periodicity, it is now natural to look
from this perspective at the most extended period of the recorded stock
market activity as dated since 1800~\cite{globalfin}. A nearly optimal
corresponding log-periodic representation versus the S$\&$P500 data is
shown in Fig.~\ref{fig:4} using the usual $\lambda = 2$. It well
reproduces the two obvious dips of the 1930s and late 1970s, and even the
broad one in the mid of the 19th century, and it also points to the one that
started in September 2000, as discussed above, as another of the same
order. The year 2002 made already clear that it can be considered as such.
The significance of this last draw down indicates that it may not fully
recover before 2004. A more detailed likely intermediate development can
be estimated by extrapolating a decelerating structure of
Fig.~\ref{fig:1}b, which allows some vital increase starting late in 2002,
possibly accompanied by accelerating log-periodic sub-patterns on
smaller time scales~\cite{Drozdz}. It, however, also indicates that in the
year 2010 the S$\&$P500 is very likely to assume values factor of a few
larger than in 2002. Extrapolating this development even further ahead in
time, one also sees that it tends around 2025 to a much larger decline 
than anything we have experienced so far. That such a scenario deserves to
be seriously taken into account we also conclude from the fact that
we had it provisionally already in 1999 at the time of extreme euphoria,
and it was signaling a large dip exactly at the begining of 21st
century, which indeed occurred.

\section{Summary}

The analysis presented above provides not only further arguments in favour
of the existence of the log-periodic component in financial dynamics,
self-similarly on various time scales, but also indicates that the
corresponding central parameter - the preferred scaling factor - may very
well be a constant close to 2. In this way it is possible to obtain a
consistent relation between the patterns and it allows more reliable
extrapolations into the future. It also allows the log-periodicity to
pretend to the status of a law. Of course, on short time scales it is a fragile
one, as the real financial market is exposed to many ``external'' factors,
such as unexpected wars or other political events, which may distort its
internal hierarchical structure on the organizational as well as on the
dynamical level. In this connection it is worth remembering that the
functional form of the log-periodic modulation so far is not supplied by
theoretical arguments and this opens room for some mathematically
unrigorous assignments of patterns, as is often needed in order to properly
interpret them. Identifying a hierarchy of time scales and a universal
preferred scaling ratio is crucial in this connection and very helpful for
real predictions. Strict fitting of the lowest order term in the Fourier
expansion of the periodic function in Eq.~(\ref{eq:logper}) is typically not
an optimal procedure. Here it serves basically as a convenient
representation to guide the eye.

\section{Acknowledgement}

We thank J. Kwapie\'n for very fruitful exchanges. S.D. acknowledges
support from Deutsche Forschungsgemeinschaft under contract Bo 56/160-1.

\end{document}